# Experimental Tailoring of a
# Three-Level $\Lambda$ System in $Tm^{3+}$:YAG

F. de Seze, A. Louchet, V. Crozatier, I. Lorgeré, F. Bretenaker, J.-L. Le Gouët

*Laboratoire Aimé Cotton, CNRS UPR 3321*

*Bâtiment 505, campus universitaire, 91405 Orsay Cedex, France*

O. Guillot-Noël, Ph. Goldner

*Laboratoire de Chimie Appliquée de l'Etat Solide, CNRS-UMR 7574,*

*ENSCP, 11 rue Pierre et Marie Curie 75231 Paris Cedex 05, France*

**Abstract** : Quantum information transfer from light to atom ensembles and *vice versa* has both basic and practical importance. Among the relevant topics let us mention entanglement and decoherence of macroscopic systems, together with applications to quantum memory for long distance quantum cryptography. Although the first experimental demonstrations have been performed in atomic vapors and clouds, rare earth ion doped crystals are also interesting media for such processes. In this paper we address $Tm^{3+}$ ions capability to behave as three-level $\Lambda$ systems, a key ingredient to convert optical excitation into a spontaneous- emission-free spin wave. Indeed $Tm^{3+}$ falls within reach of light sources that can be stabilized easily to the required degree. In the absence of zero-field hyperfine structure we apply an external magnetic field to lift the nuclear spin degeneracy in $Tm^{3+}$:YAG. We experimentally determine the gyromagnetic tensor components with the help of spectral hole-burning techniques. Then appropriate orientation of the applied field enables us to optimize the transition probability ratio along the two legs of the $\Lambda$. The resulting three-level $\Lambda$ system should suit quantum information processing requirements.

# 1 Introduction

Although quantum physics offers the paradigm for microscopic world description, there has been growing interest in observing macroscopic size quantum processes. The idea of mapping a quantum excitation onto a large ensemble of elementary systems, that can be traced back to the early days of quantum mechanics [1], has been successfully revisited in the frame of atomic physics [2]-[8]. In these works an optical collective excitation entangles the macroscopic ensemble state. The activity in the field is stimulated by the potential application to the quantum memories needed for long distance quantum cryptography. In most schemes, the optical excitation is stored as a spin wave in the atomic ensemble [9]. The basic ingredient to convert incident photons into spin excitation is known as a three-level $\Lambda$ system. The spin state is built from two nearby sublevels |1> and |2> of the atomic ground state that are connected to a common upper state |3> by optical transitions along the two legs of the $\Lambda$. This way one avoids decoherence by spontaneous emission. One recalls the stored information to life by turning back the spin coherence |1><2| into the optical coherence |1><3|.

So far, ensemble entanglement has been demonstrated only in atomic vapors [[4]-[6]] and beams [3] or in laser cooled atom clouds [2], [7], [8]. However rare earth ion doped crystals (REIC) also appear as promising candidates in the quest for macroscopic quantum effects. They offer properties similar to atomic vapors with the advantage of no atomic diffusion. At low temperature (< 4K) the optical coherence lifetime may reach several ms in these materials and a hyperfine coherence lifetime of tens of seconds has been reported in $Pr^{3+}$: $Y_2SiO_5$ [10], [10]. Given the absence of atomic motion, extremely long population lifetime can be observed. Storage of light, with quantum memory in prospect, has given rise to Electromagnetically Induced Transparency (EIT) investigations [12], [13]. A storage time greater than one second has been observed in the most recent work [13]. Among all the rare earth ions, non-Kramers ions with an even number of $4f$ electrons present the longest coherence lifetime [14]. With the additional condition of a few tens of MHz hyperfine splitting in the electronic ground state, one is practically left only with $Eu^{3+}$ and $Pr^{3+}$ compounds as good rare earth ion candidates for $\Lambda$ systems. However, only dye lasers are available at operation wavelengths in $Eu^{3+}$: $Y_2SiO_5$ (580nm) and $Pr^{3+}$: $Y_2SiO_5$ (606nm). Because of the high frequency noise generated by the dye jet, this is a challenging task to reach the sub-kHz line width and jitter that can match the long optical coherence lifetime offered by REIC. Actually few dye laser systems throughout the world offer such a high degree of stability [13], [15], [16], [17].

Some other non-Kramers REIC fall within reach of more tractable lasers. For instance, the $^3H_6(0) \rightarrow {}^3H_4(0)$ transition of $Tm^{3+}$ in $Y_3Al_5O_{12}$ (YAG) can be driven by semiconductor lasers. Such lasers can be stabilized easily to sub-kHz linewidth and jitter [18]. The $Tm^{3+}$: YAG compound has been widely used for coherent transient-based signal-processing applications [19], [20]. The atomic coherence associated with the $^3H_6(0) \rightarrow {}^3H_4(0)$ transition at 793nm exhibits a lifetime of 70µs that has been observed to grow to 105 µs under moderate magnetic field [21]. Although $Tm^{3+}$: YAG exhibits no hyperfine structure, Thulium possesses a I= ½ nuclear spin. Lifting the nuclear spin degeneracy with an external magnetic field may offer a way for building a $\Lambda$ system, provided one is able to relax the nuclear spin $\Delta M_I = 0$ selection rule. Indeed electronic excitation cannot flip the nuclear spin and this can forbid optical transition along one $\Lambda$ leg. Observation of long lifetime spectral hole burning in $Tm^{3+}$: YAG under applied magnetic field proved the nuclear spin selection rule can indeed be relaxed [22] but the optical transition probability ratio was not measured. In recent papers [23], [24] we theoretically investigated the nuclear state mixing induced by an external magnetic field and determined the best field orientation for optimizing the relative strength of the optical transitions along the two $\Lambda$ legs. In the present paper we measure the transition probability ratio and show that $Tm^{3+}$: YAG can actually operate as a $\Lambda$ system.

The paper is arranged as follows. Section 2 sets the theoretical framework. Relying on $Tm^{3+}$: YAG symmetry properties, we propose an effective magnetic field geometrical model to describe the level mixing effect and demonstrate the connection between the transition probability ratio and the relative level splitting in upper and lower electronic state. The site diversity is taken into account. Section 3 is devoted to the experimental characterization of the $\Lambda$ system. The relevant parameters are deduced from spectral hole-burning spectra, the experiment conditions being specified by the magnetic field orientation and the light beam direction of polarization. We conclude in section 4.

## 2 Theoretical framework

### 2.1 Non-Kramers ions and low symmetry sites

With an even number of electrons, $Tm^{3+}$ is a non-Kramers ion whose electronic level degeneracy is totally lifted by the crystal field at sites with less than threefold rotational symmetry [25]. In such low symmetry substitution sites, the resulting singlet states, typically spaced by energy intervals of several hundreds of GHz, are little affected by spin fluctuation. At low temperature, such ions lying in the lower energy level of Stark multiplets then behave in a way similar to atoms in a low pressure vapor. They can be prepared in long lifetime superposition states.

The only natural isotope of Thulium ($^{169}Tm$) exhibits a ½ nuclear spin. In low symmetry sites the two nuclear states are degenerate in the absence of external magnetic field. This is connected with time reversal symmetry [25]. The electronic singlet states are time reversal eigenstates, and the total electronic angular momentum $\vec{J}$ is an odd operator with respect to time-reversal. Therefore the quantum expectation value $<\vec{J}>$ vanishes in any Stark sub-level, which is known as spin quenching. Hence, interactions expressed in terms of $\vec{J}$, such as hyperfine coupling, also vanish to first order of perturbation. In $Tm^{3+}$ the hyperfine coupling also fails to lift nuclear spin degeneracy to second order because of the ½ nuclear spin value [26].

### 2.2 Building a three-level system in $Tm^{3+}$

The nuclear degeneracy reflects the fact that the nuclear spin commutes with the electronic Hamiltonian. As a result the two nuclear spin substates in the ground state $^3H_6(0)$ do not compose a $\Lambda$ three level system with a single common nuclear spin substate in upper level $^3H_4(0)$. Indeed optical excitation cannot flip the nuclear spin. Interaction with an external magnetic field $\vec{B}$, if restricted to the nuclear Zeeman effect, does lift the nuclear spin degeneracy but is unable to overcome this selection rule. Fortunately, the cross-coupling of electronic Zeeman effect and hyperfine interaction occurs to second order of perturbation and provides us with the needed nuclear spin state mixing. Let the nuclear Zeeman, the electronic Zeeman and the hyperfine Hamiltonians be respectively given by the following expressions:

$$H_{nZ} = -g_n \beta_n \vec{B}.\vec{I} \ , \ H_{eZ} = -g \beta \vec{B}.\vec{J} \ , \ H_{hyp.} = A \vec{I}.\vec{J} \tag{1}$$

where $A$ represents the hyperfine coupling constant, $g_n$ and $g$, $\beta_n$ and $\beta$ respectively stand for the nuclear and electronic Lande factors and for the nuclear and Bohr magnetons. Then, in the basis of the zero-magnetic field eigenstates, to second order of perturbation, the nuclear Zeeman effect, as exacerbated by cross electronic Zeeman and hyperfine interactions, can be expressed by the effective Hamiltonian [26]:

$$H'_{nZ} = \sum_{i,j=x,y,z} \gamma_{ij} B_i I_j \tag{2}$$

where:

$$\gamma_{ij} = -g_n \beta_n - g \beta \Lambda_{ij} \tag{3}$$

with:

$$\Lambda_{ij} = A \sum_{n \neq 0} \frac{\langle 0 | J_i | n \rangle \langle n | J_j | 0 \rangle}{E_n - E_0} \tag{4}$$

The lower energy substate of the electronic Stark multiplet is denoted $|0\rangle$ and the sum runs over the other multiplet sublevels $|n\rangle$. Sites with $D_2$ symmetry are specially attractive. Indeed the tensor $\Lambda_{ij}$ is diagonal in the site frame that is built along the three orthogonal two-fold axes of symmetry $Ox$, $Oy$ and $Oz$ [27]. Indeed, $J_x$, $J_y$ and $J_z$ operators transform as the electric dipole operators $x$, $y$, $z$. In a $D_2$ point symmetry, the electric dipole transitions are only allowed along the $Ox$, the $Oy$ or the $Oz$ direction. The $\Lambda_{ij}$ components with $i \neq j$ thus vanish and the tensor is diagonal in the site frame. Therefore the effective nuclear Zeeman Hamiltonian reduces to:

$$H'_{nZ} = \gamma_x B_x I_x + \gamma_y B_y I_y + \gamma_z B_z I_z \qquad (5)$$

where

$$\gamma_x = \gamma_{xx}, \gamma_y = \gamma_{yy}, \gamma_z = \gamma_{zz} \qquad (6)$$

In ground and excited states the gyromagnetic factors take on different values. They are respectively denoted as $\gamma_i^{(g)}$ and $\gamma_i^{(e)}$. These parameters have been determined theoretically for a $Tm^{3+}$ ion doped YAG matrix that offers $D_2$ symmetry substitutions sites [24]. The results can be summarized qualitatively in the following way:

i)   the gyromagnetic tensor is strongly anisotropic in both ground and upper levels, $^3H_6(0)$ and $^3H_4(0)$, i.e. $\gamma_y^{(e)} >> \gamma_x^{(e)}, \gamma_z^{(e)}$ and $\gamma_y^{(g)} >> \gamma_x^{(g)}, \gamma_z^{(g)}$

ii)  the anisotropy in the (x, z) plane is similar in ground and upper levels, i.e. $\left( \gamma_x^{(e)} / \gamma_z^{(e)} - \gamma_x^{(g)} / \gamma_z^{(g)} \right)^2 << \left( \gamma_x^{(e)} / \gamma_z^{(e)} + \gamma_x^{(g)} / \gamma_z^{(g)} \right)^2$

iii) the (x, y) anisotropy is much larger in ground state than in upper state, i.e. $\gamma_y^{(e)} / \gamma_x^{(e)} << \gamma_y^{(g)} / \gamma_x^{(g)}$

An effective magnetic field picture helps to predict the magnetic field orientation that optimizes the balance of the transition probabilities along the two branches of the Λ.

## 2.3   Effective magnetic field picture

In either ground or upper level, according to Eq. (5), the frequency splitting reads as:

$$\Delta = \left[ \gamma_x^2 B_x^2 + \gamma_y^2 B_y^2 + \gamma_z^2 B_z^2 \right]^{1/2} \qquad (7)$$

for ½ spin states. Let us define an effective field unit vector as:

$$\hat{B}_{eff} = \left( \gamma_x B_x / \Delta, \gamma_y B_y / \Delta, \gamma_z B_z / \Delta \right) = (X, Y, Z) \qquad (8)$$

In terms of this unit vector the effective nuclear Zeeman Hamiltonian, as defined in Eq. (5), turns into:

$$H'_{nZ} = \left( \hat{B}_{eff} \cdot \vec{I} \right) \Delta \qquad (9)$$

In either electronic level the wavefunctions can be expressed as the product of a common electronic part $|\psi_{el}\rangle$ and of the $H'_{nZ}$ eigenvectors. Indeed $H'_{nZ}$, of order $<10^{-2}$cm$^{-1}$, can be regarded as a perturbation of the electronic Hamiltonian, of order 10 to 100 cm$^{-1}$. The $H'_{nZ}$ eigenvectors read as $|1\rangle = a_1 |+\rangle + b_1 |-\rangle$ or $|2\rangle = a_2 |+\rangle + b_2 |-\rangle$, where $|+\rangle$ and $|-\rangle$ stand for $I_z$ eigenvectors and the coefficients are given by:

$$a_1 = \frac{1}{\sqrt{2}} \frac{X + iY}{(1+Z)^{1/2}}, \quad b_1 = \frac{1}{\sqrt{2}} \frac{1+Z}{(1+Z)^{1/2}} \qquad (10)$$

$$a_2 = \frac{1}{\sqrt{2}} \frac{X + iY}{(1-Z)^{1/2}}, \quad b_2 = -\frac{1}{\sqrt{2}} \frac{1-Z}{(1-Z)^{1/2}} \qquad (11)$$

The ratio of the transition probabilities along the two branches of the $\Lambda$, or in other words the branching ratio to the forbidden transition, is then easily expressed as:

$$R = \frac{|<2\,|\,3>|^2}{|<1\,|\,3>|^2} = \frac{1 - \cos \alpha^{(eff)}}{1 + \cos \alpha^{(eff)}} = \frac{\left\| \hat{B}_{eff}^{(e)} \times \hat{B}_{eff}^{(g)} \right\|^2}{\left( 1 + \hat{B}_{eff}^{(e)} \cdot \hat{B}_{eff}^{(g)} \right)^2} \qquad (12)$$

where $\alpha^{(eff)}$ represents the angle of the effective field directions in lower and upper electronic levels. Therefore the branching ratio R vanishes when $\hat{B}_{eff}^{(e)}$ is parallel to $\hat{B}_{eff}^{(g)}$ and equals unity when $\hat{B}_{eff}^{(e)}$ is orthogonal to $\hat{B}_{eff}^{(g)}$.

## 2.4  Optimizing the magnetic field orientation

Theory predicts $\gamma_y^{(e)} >> \gamma_x^{(e)}, \gamma_z^{(e)}$ and $\gamma_y^{(g)} >> \gamma_x^{(g)}, \gamma_z^{(g)}$. Because of this strong anisotropy, should the applied magnetic field have a significant component along $Oy$, the effective field in both states would be aligned along $Oy$, resulting in a very small relative angle $\alpha^{(eff)}$ of the effective fields in both states and a small branching ratio. Thus an optimally oriented field shall be strongly slanted with respect to $Oy$. However $B$ should not be orthogonal to $Oy$.

Indeed, according to theory, $\gamma_x^{(e)}/\gamma_z^{(e)} \simeq \gamma_x^{(g)}/\gamma_z^{(g)}$, which means that upper and lower level effective field components are nearly collinear in plane $xOz$. Therefore the branching ratio is expected to be small when $B_y = 0$. Theory also predicts that anisotropy in ground state is much larger than in excited state. Let us start with an applied field lying in the $xOz$ plane and let us add some field component along $Oy$. Because of the much larger anisotropy in the ground state the *effective* field component along $Oy$ will grow much faster in ground state than in upper state, which will result in a large *effective* field relative angle at small *applied* field component along $Oy$.

To be more specific, let us suppose that the applied field is initially directed along $Ox$ and we look for optimal direction within the plane $xOy$. From previous discussion we expect the maximum branching ratio will be obtained by slightly tilting the applied field away from direction $Ox$ by an angle $\theta$. The applied field component along $Oy$ is given by $B\theta$ where $\theta \ll 1$. The effective field unit vectors read as:

$$\hat{B}_{eff}^{(j)} = \left( r_j, \theta, 0 \right) / \sqrt{r_j^2 + \theta^2} \qquad (13)$$

where $r_j = \gamma_x^{(j)}/\gamma_y^{(j)}$ and $(j)$ stands for the level label *(e)* or *(g)*. The cross product reads as:

$$\left\| \hat{B}_{eff}^{(e)} \times \hat{B}_{eff}^{(g)} \right\| = \sin \alpha_{eff} = \frac{\left| \left( r_e - r_g \right) \theta \right|}{\sqrt{r_e^2 + \theta^2} \sqrt{r_g^2 + \theta^2}} \qquad (14)$$

and is maximum at the following optimal tilt angle value:

$$\theta_0 = \sqrt{r_e r_g} \qquad (15)$$

The maximum cross product and branching ratio maximum values respectively read as:

$$\left\| \hat{B}_{eff}^{(e)} \times \hat{B}_{eff}^{(g)} \right\|_{max} = \frac{\left| \left( r_e - r_g \right) \right|}{\left( r_e + r_g \right)} \qquad (16)$$

$$R_{\max} = \left( \frac{\sqrt{r_e} - \sqrt{r_g}}{\sqrt{r_e} + \sqrt{r_g}} \right)^2 \qquad (17)$$

According to the definition of level splitting in Eq. (7), the maximum branching ratio can be written as:

$$R_{\max} = \left( \frac{\sqrt{\Delta_g / \gamma_y^{(g)}} - \sqrt{\Delta_e / \gamma_y^{(e)}}}{\sqrt{\Delta_g / \gamma_y^{(g)}} + \sqrt{\Delta_e / \gamma_y^{(e)}}} \right)^2 \qquad (18)$$

where $\Delta_g$ and $\Delta_e$ represent the ground and excited state level splitting *at $B_y$ =0.* In terms of the level splitting and of $\gamma_y$ Eq.(15) reads as:

$$\theta_0 = \sqrt{\Delta_g \Delta_e} / \sqrt{\gamma_y^{(g)} \gamma_y^{(e)}} \qquad (19)$$

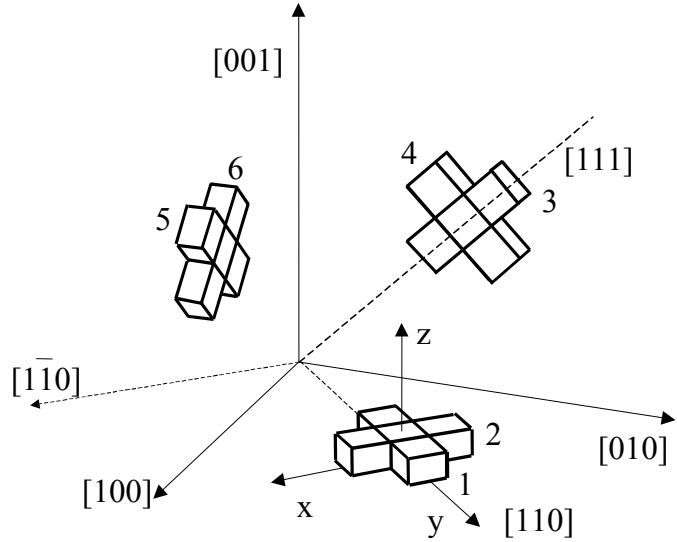

Figure 1 Tm$^{3+}$ substitution sites in the Yttrium Aluminium garnet (YAG) matrix. The crystal cell axes are denoted [100], [010] and [001]. There are six cristallographically equivalent, orientationally inequivalent sites, each of them being defined by a local frame. The local axes *x, y,* and *z* for site 1 are represented in the figure. The *Oy* axis of site 1 coincides with the crystal cell diagonal [110]. In each site, the transition dipole moment, represented by an oblong, is directed along the local *Oy* axis. Experiments were carried out with the light passing along the [1 $\overline{1}$0] axis and the light polarization was directed along the cell diagonal [111].

In the Appendix we show that these results can be generalized to situations where B is not contained in plane *xOy*. We show that, whatever the orientation of the applied magnetic field projection in plane *xOz*, a *lower boundary* to the optimal branching ratio R$_{\max}$ and the optimal direction with respect to $Oy$ are still given by Eqs. (18) and (19) where the splitting values $\Delta_g$

and $\Delta_e$ are measured *at $B_y = 0$.*

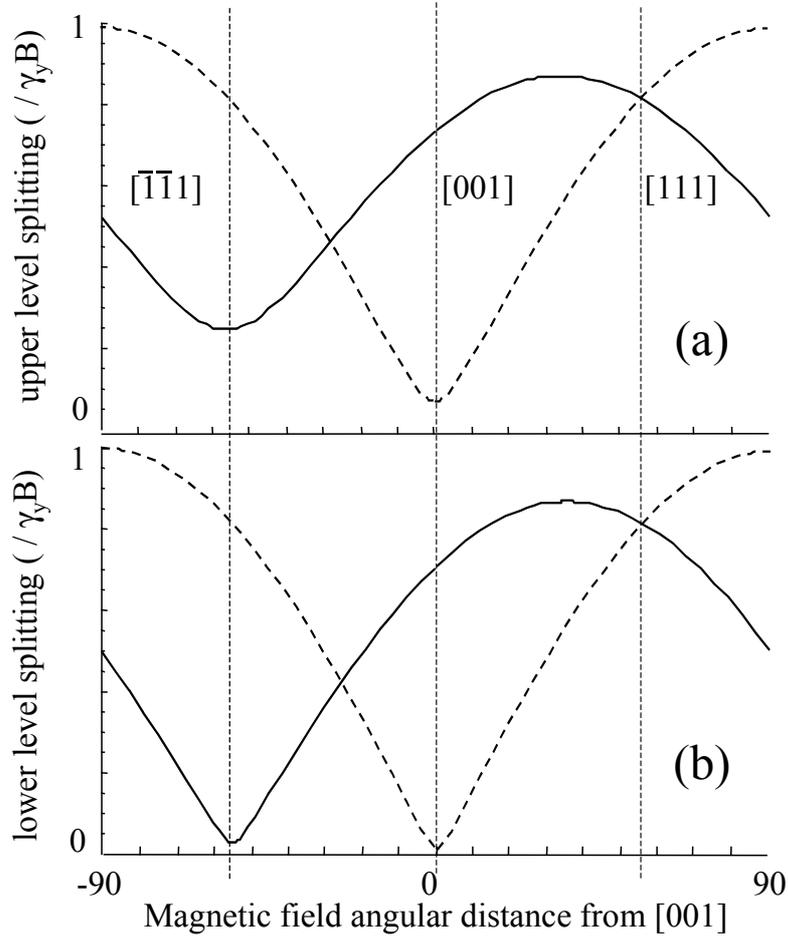

Figure 2 theoretical prediction for splitting in upper (a) and lower (b) electronic levels for site 1 (dashed line) and 3 and 5 (solid line) as a function of the magnetic field orientation. The field lies within the $(1\overline{1}0)$ bisector plane of the crystal cell and its direction in this plane is defined by the angular distance from vertical axis [001].

## 2.5 Crystal symmetry and site selection

So far we discussed the external magnetic field orientation in a site local frame. In YAG the Thulium dopant ion substitutes for Yttrium in six different crystallographically equivalent and orientationally inequivalent sites (see Figure 1). As a consequence of the $D_2$ site symmetry, the optical transition dipole moment in $Tm^{3+}$ can only be oriented along a site axis. More specifically the electronic wave function distortion by the crystal field simultaneously affects the gyromagnetic tensor and the optical transition dipole moment that turns out to be directed along the $Oy$ local axis [28].

The fixed dipole direction offers a convenient way to select specific sites. Let the light beams propagate along $[1\bar{1}0]$, with polarization directed along the cell diagonal [111]. Then the light beams are cross-polarized with sites 2, 4 and 6 and do not interact with them. In addition, dipoles in sites 1, 3 and 5 are slanted at the same angle from the polarization direction. They interact with the light beams with the same Rabi frequency and contribute the same amount to the crystal optical density. Finally, the sites 3 and 5 are magnetically equivalent if the applied magnetic field is orthogonal to $[1\bar{1}0]$, lying in the $(1\bar{1}0)$ bisector plane of the crystal cell. Hence one is left with only two site classes, that are respectively comprised of site 1 alone and of sites 3 and 5.

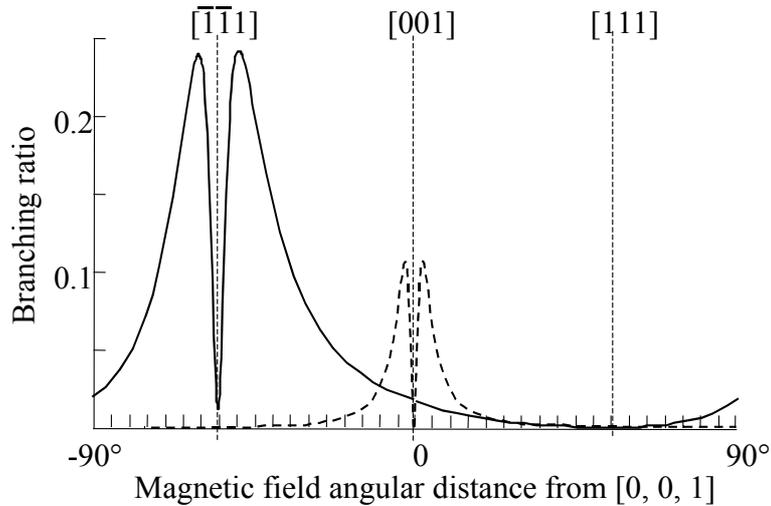

Figure 3 theoretical prediction for the $\Lambda$ system branching ratio in site 1 (dashed line) and 3 and 5 (solid line) as a function of the magnetic field orientation. The field lies within the $(1\bar{1}0)$ bisector plane of the crystal cell and its direction in this plane is defined by the angular distance from vertical axis [001].

The theoretical prediction for the splitting in lower and upper electronic level [24] is displayed in Figure 2. The ground state splitting in sites 3 and 5 (respectively, in site 1) exhibits deep minima in direction $[\bar{1}\bar{1}1]$ (respectively [001]) that corresponds to vanishing $B_y$ component. This reflects the gyromagnetic tensor anisotropy and the relative large size of $\gamma_y$. As discussed in section 2.4, and illustrated in Figure 3, the optimal branching ratio value is observed in the vicinity of these orientations. The small value of the branching ratio in direction $[\bar{1}\,\bar{1}1]$ for sites 3 and 5 corresponds to the small anisotropy disparity in the plane $xOz$ of these sites' local frame, which results in a small angle between the effective magnetic field vectors in upper and lower electronic states. In site 1, the branching ratio vanishes in direction [001]. Indeed the applied magnetic field is then directed along the local frame axis $Oz$, together with the effective fields in both levels, which makes the effective field angle vanish.

Although small, the anisotropy disparity in plane $xOz$ is large enough to induce significant level splitting differences between site 1 and sites 3 and 5 at $B_y = 0$. In site 1 (respectively 3 and 5), when $B_y$ vanishes as the applied field is directed along [001] (respectively $[\bar{1}\,\bar{1}1]$), the local frame components of $B$ read as $(0, 0, B)$ (respectively $\left(B\sqrt{2/3}, 0, B/\sqrt{3}\right)$). The component along $Ox$ makes the difference since, according to theory [24], $\gamma_x^{(e)}$ differs more from $\gamma_x^{(g)}$ than $\gamma_z^{(e)}$ from $\gamma_z^{(g)}$. As a consequence, the upper and lower state splitting ratio is larger in sites 3 and 5 in direction $[\bar{1}\,\bar{1}1]$ than in site 1 in direction [001].

In accordance with discussion in section 2.4, smaller splitting ratio in site 1 also results in smaller optimal branching ratio. The optimal branching ratio in site 1, when $B$ is slightly tilted from direction [001], is exactly expressed by Eq. (18) since $B$ is then contained in plane $yOz$. As discussed in the Appendix, this equation only gives a lower boundary of the optimal branching ratio in sites 3 and 5 when $B$ is slightly tilted from direction $[\bar{1}\,\bar{1}1]$. However the exact value 0.236 appears to be very close to the approximate value 0.235, given by Eq. (18). The field $B$ lying in the $(1\bar{1}0)$ bisector plane at angle $\Theta$ from vertical axis [001], its $B_y$ component in the local frames of sites 3 and 5 reads as:

$$B_y / B = \frac{1}{2}\sin\Theta + \frac{1}{\sqrt{2}}\cos\Theta = \frac{\sqrt{3}}{2}\sin\left(\Theta - \Theta_{[\bar{1}\,\bar{1}1]}\right) \qquad (20)$$

where $\Theta_{[\bar{1}\bar{1}1]}$ corresponds to direction $[\bar{1}\bar{1}1]$. To relate the optimal field orientation in the crystal frame to the local frame tilt angle $\theta_0$, as calculated in the Appendix, one identifies $B_y$ with $B\sin\theta_0$. It results that optimal field direction heads at angle $\delta\Theta_0 = 2\theta_0/\sqrt{3}$ from $[\bar{1}\bar{1}1]$ in the crystal frame.

Let us concentrate on this field configuration that optimizes the branching ratio in sites 3 and 5. The light beams, polarized along [111], only interact with sites 1, 3 and 5, the three of them contributing an equal amount to absorption. Therefore the optimized sites 3 and 5 represent two thirds of the optical density. Since the magnetic field is close to direction $[\bar{1}\bar{1}1]$, the splitting is much larger in site 1 than in sites 3 and 5, as illustrated in Figure 2. This will help to remove site 1 ions from the spectral region of investigation.

Other polarization selections lead to less efficient site selection, the optimized ions contributing a lower amount to the optical density. For instance, this is so when light is polarized along the vertical axis [001]. Then light is cross-polarized with sites 1 and 2 and interacts with sites 3, 4, 5 and 6 on an equal footing. Since the three level system is optimized in sites 3 and 5, only half the available optical density can be used for three level operation.

In summary, directing the applied magnetic field at angle $\pm\delta\Theta_0$ from $[\bar{1}\bar{1}1]$ in the $(1\bar{1}0)$ bisector plane of the crystal cell and polarizing the light field along the cell diagonal [111], one should be able to select a single class of ions, offering an optimized branching ratio and representing 2/3 of the available optical density. The tilt angle $\delta\Theta_0$ and the corresponding optimal branching ratio have to be determined experimentally. The next section is devoted to the relevant measurements.

# 3  Experimental optimization of the Λ system

In order to proceed with the evaluation of the three level system we first have to determine the optimal orientation of the magnetic field and the corresponding optimal value of the branching ratio $R$. According to Eqs. (7), (8) (12) and (15), all the relevant quantities can be expressed in terms of the gyromagnetic tensor coefficients $\gamma_i^{(g)}$ and $\gamma_i^{(e)}$. Instead of measuring the branching ratio directly, it is easier to deduce the $\gamma_i^{(g)}$ and $\gamma_i^{(e)}$'s from level splitting measurements, and then to derive the optimal magnetic field orientation and branching ratio from the experimental $\gamma_i^{(g)}$ and $\gamma_i^{(e)}$ values. We resort to hole burning spectroscopy in order to measure the level splitting.

## 3.1  Spectral hole burning mechanism

Spectral hole burning occurs when an inhomogeneously broadened absorption line is irradiated by a monochromatic laser. When nuclear spin degeneracy is lifted by the applied magnetic field, a monochromatic burning laser simultaneously excites ions along four different transitions, at given frequency $\omega_0$. The different burning and probing schemes are summarized in Figure 4. We have to consider two different burning schemes, depending whether the burning transition is allowed or not. Initially the two ground state sublevels are evenly populated, given the energy spacing is always much smaller than thermal energy at working temperature. Let us first consider that the burning beam excites an allowed transition. The monochromatic burning laser standing at frequency $\omega_0$, the ions that are resonant at $\omega$ undergo excitation to upper level at rate $P(\omega - \omega_0)$. The function $P(\omega - \omega_0)$ is maximum at 0 and is proportional to the homogeneous line shape. The decay to the same sublevel is an allowed process, proceeding at rate $\Gamma$. Decay to the other ground state sublevel is partly forbidden. We denote the corresponding decay rate as $\Gamma R_1$, where the branching ratio $R_1$ may differ from the optical transition branching ratio $R$. Indeed decay proceeds mainly through the shelving state $^3F_4$ and may obey specific selection rules. As a result a fraction $R_1/(1+R_1)$ of the excited ions are pumped to the second ground state sublevel. Therefore the pumping rate reads as $P(\omega-\omega_0)R_1/(1+R_1)$. When burning takes place on a forbidden transition, excitation to the upper level is achieved at rate $P(\omega-\omega_0)R$. A fraction $1/(1+R_1)$ of the ions decays to the other sublevel. The pumping rate is $P(\omega-\omega_0)R/(1+R_1)$.

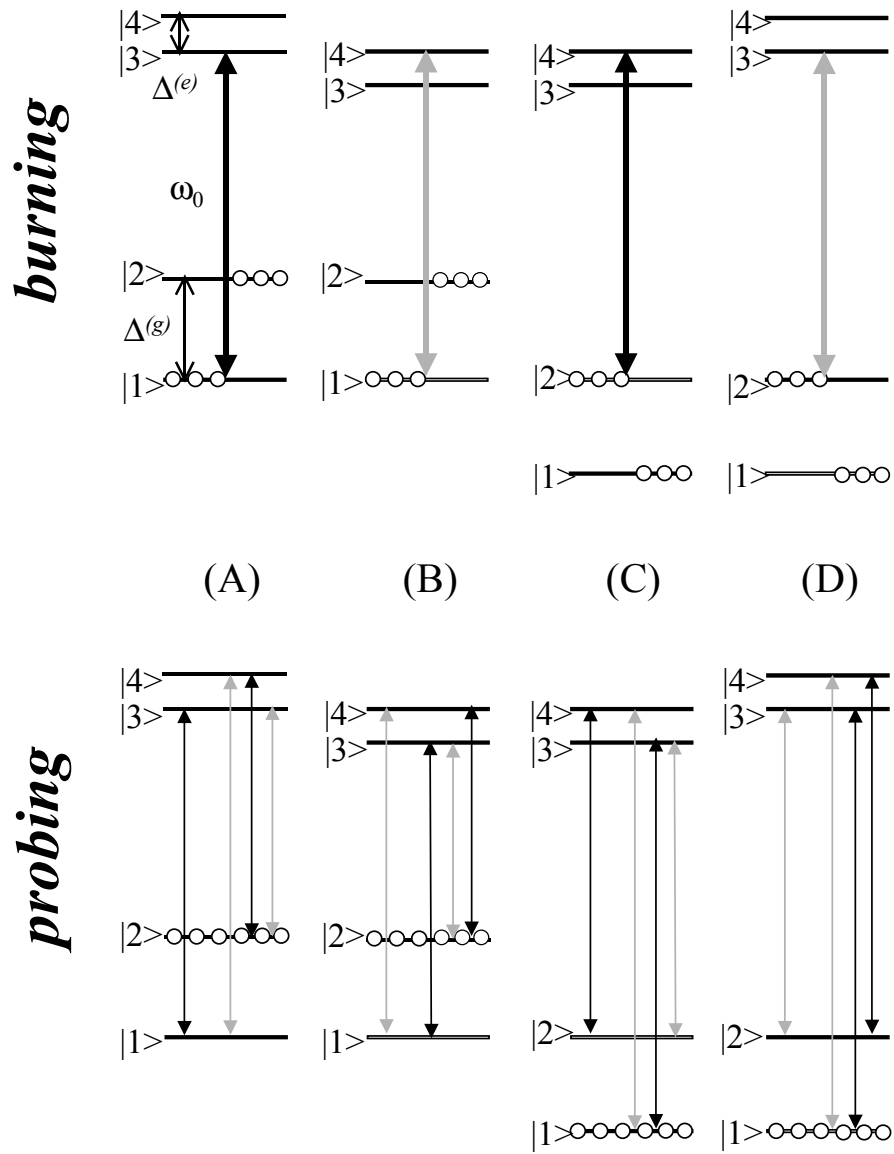

Figure 4 spectral hole burning scheme under excitation by a magnetic field. Four different transitions are simultaneously excited at the pumping frequency $\omega_0$ Allowed and forbidden transitions are depicted by black and grey arrows respectively. Initially ions, represented by open circles, are evenly distributed over the ground state sublevels. Exposure at $\omega_0$ results in optical pumping to the off-resonance sublevel. At probing, absorption is reduced (respectively increased) for lines connected to the depleted (respectively populated) sublevels.

The optical pumping between ground state sublevels conflicts with relaxation processes that tend to restore thermal equilibrium at rate $\Gamma_0 \ll \Gamma$. In order to avoid saturation broadening of the depleted spectral region, one has to take care that $P(\omega - \omega_0) R_1 /(1 + R_1) < \Gamma_0$ and $P(\omega - \omega_0) R /(1 + R_1) < \Gamma_0$. When this condition is satisfied, the inhomogeneously broadened absorption profile exhibits an homogeneously broadened hole at $\omega_0$. When $P(0) R_1 /(1 + R_1) \gg \Gamma_0$ the sample is totally bleached at $\omega_0$ and the transparent spectral window covers an interval much broader than the homogeneous profile.

After burning at $\omega_0$, the absorption spectrum exhibits additional features. Indeed, while the resonantly excited sublevel is depleted, the off-resonant sublevel population is increased, which gives rise to anti-holes located at $\omega_0 \pm \Delta^{(g)}$ where $\Delta^{(g)}$ denotes the ground state splitting. One has also to take account of the upper level splitting $\Delta^{(e)}$. Transitions from the depleted ground state sublevel give rise to holes not only at $\omega_0$ but also at frequency $\omega_0 \pm \Delta^{(e)}$. In the same way, anti-holes occur not only at $\omega_0 \pm \Delta^{(g)}$ but also at $\omega_0 \pm \left( \Delta^{(g)} - \Delta^{(e)} \right)$ and $\omega_0 \pm \left( \Delta^{(g)} + \Delta^{(e)} \right)$.

In the prospect of data analysis it is worth sorting holes and anti-holes in terms of allowed and forbidden transitions, considering both the burning and probing stages. Allowed and forbidden transitions contribute at $\omega_0$, $\omega_0 \pm \Delta^{(e)}$, and $\omega_0 \pm \Delta^{(g)}$. Only allowed (respectively forbidden) transitions contribute at $\omega_0 \pm \left( \Delta^{(g)} - \Delta^{(e)} \right)$ (respectively $\omega_0 \pm \left( \Delta^{(g)} + \Delta^{(e)} \right)$). The different transition probabilities are detailed for each feature in Table 1. Contributions with allowed probing transitions shall be easier to observe. At $\omega_0$ and $\omega_0 \pm \left( \Delta^{(g)} - \Delta^{(e)} \right)$ (respectively $\omega_0 \pm \Delta^{(e)}$ and $\omega_0 \pm \Delta^{(g)}$), the relevant contributions result from burning along an allowed (respectively a forbidden) transition, with a pumping rate $P(\omega-\omega_0)R_1/(1+R_1)$ (respectively $P(\omega-\omega_0)R/(1+R_1)$).

| Hole or anti-hole position | $\omega_0$ | | $\omega_0\pm\Delta^{(e)}$ | | $\omega_0\pm\left(\Delta^{(g)}-\Delta^{(e)}\right)$ | $\omega_0\pm\Delta^{(g)}$ | | $\omega_0\pm\left(\Delta^{(g)}+\Delta^{(e)}\right)$ |
|---|---|---|---|---|---|---|---|---|
| Burning transition | A | F | A | F | A | A | F | F |
| Probing transition | A | F | F | A | A | F | A | F |

Table 1 relative transition probability for the different hole and anti-hole burning and probing. The forbidden (F) and allowed (A) transitions probabilities are related by F= R A. Each column corresponds to one specific contribution to the observed feature (see Figure 4). Contributions with forbidden probing transitions (hatched columns) are not expected to be observed outside the very field direction that optimizes the branching ratio R.

## 3.2  Experimental setup

The burning and probe light is provided by a fixed frequency single mode stabilized extended cavity semiconductor laser [29] operating at 793nm. Burning and probing sequences are shaped by an external acousto-optic shifter that also performs the required frequency scan for probing the absorption profile. The 5mm-thick crystal composition is 0.1 at. % $Tm^{3+}$:YAG. The optical density at 793nm is 0.3. The crystal is cut perpendicular to direction $[1\bar{1}0]$ along which the light beam propagates. The magnetic field, generated by a pair of Samarium-Cobalt button magnets, is made orthogonal to $[1\bar{1}0]$. Depending on the magnet size, an induction as high as 0.45T can be obtained. The sample, together with the magnets, is positioned inside a liquid helium bath cryostat. The laser is focused on the sample to a 1/e² diameter of ~800μm. Burning is achieved by a sequence of 10 shots of 50μs each, evenly spaced by 10ms-intervals. The peak power amounts to ~1mW. During the probing stage, the AO shifter attenuates the light beam by a factor of 100 and scans the frequency over a ~20MHz wide interval in 750μs. The transmitted probe light is detected on an avalanche photodiode (HAMAMATSU C5460). We repeat the whole burning and probing sequence at rate 6s[-1] and we wait until stationary regime is reached before recording the transmission spectra with a digital oscilloscope TEKTRONIX TDS3032B. Keep in mind that the features conventionally referred to as "holes" and "anti-holes" will appear this way when observed in absorption profiles. We shall keep this naming although they manifest themselves just the opposite way in the experimental transmission spectra we display in this work.

### 3.3 Measurement of energy level splitting at $B_y = 0$

In site 1 (respectively 3 and 5), when $B_y$ vanishes as the applied field is directed along [001] (respectively $[\overline{1}\,\overline{1}1]$), the level splitting reads as $B\gamma_z$ (respectively $B\sqrt{\left(2\gamma_x^2+\gamma_z^2\right)/3}$). Therefore, combining measurements in these two applied field directions, one should be able to determine $\gamma_x$ and $\gamma_z$. However one must remember that the branching ratio to the forbidden transition strictly vanishes in site 1 in direction [001] since the applied field is then directed along axis $Oz$ in the local frame. Since the Zeeman Hamiltonian reads as $\gamma_z B_0 I_z$ the nuclear spin projection along $Oz$ is a good quantum number and no nuclear spin flip can be induced by an optical transition. Therefore hole burning does not operate in site 1 in that field direction. Slightly away from that direction the branching ratio increases, as illustrated in Figure 3, but the ground state level splitting also grows rapidly as can be seen in Figure 2. With our present equipment the field orientation is not defined with enough precision and we must confine ourselves to the site 3 and 5 splitting measurement in direction $[\overline{1}\,\overline{1}1]$. Then results are less sensitive to field orientation since this corresponds to a splitting minimum. As discussed in section 2.4, measuring the level splitting and $\gamma_y$ is enough to get a lower boundary to the branching ratio R.

In the experiment the light beam, polarized along direction [111], interacts with ions in sites 1, 3 and 5. The 0.45T-magnetic field is directed along $[\overline{1}\,\overline{1}1]$ so that $B_y$ vanishes in sites 3 and 5. The sample is cooled by a flow of cold helium gas but is not immersed in liquid helium. By adjusting the sample temperature we tune the ground state relaxation rate $\Gamma_0$ in order to avoid saturation broadening of the holes and anti-holes. As discussed in section 3.1 one has to make $\Gamma_0$ larger than the optical pumping rate. The transmission spectrum in Figure 5 represents a 750μs-long scan over a 19MHz-wide interval after a 10ms waiting time, given the 10ms lifetime of the shelving state $^3F_4$. Burning is observed at the central position A. The hole C is ascribed to a transition to the upper level and should be located at distance $\Delta_e$ from the burning position. The antiholes B and D are expected to be located at $\left(\Delta_g - \Delta_e\right)$ and $\Delta_g$ respectively from the burning position. From the data we deduce the following splitting values:

$$\Delta_e^{[\bar{1}\bar{1}1]} = 14.4 \pm 0.1\, MHz\,/\,Tesla \;,\; \Delta_g^{[\bar{1}\bar{1}1]} = 15.3 \pm 0.1\, MHz\,/\,Tesla \tag{21}$$

where the connection with $\gamma_x$ and $\gamma_z$ can be expressed as:

$$\Delta^{[\bar{1}\bar{1}1]} = \sqrt{\left(2\gamma_x^2 + \gamma_z^2\right)/3} \tag{22}$$

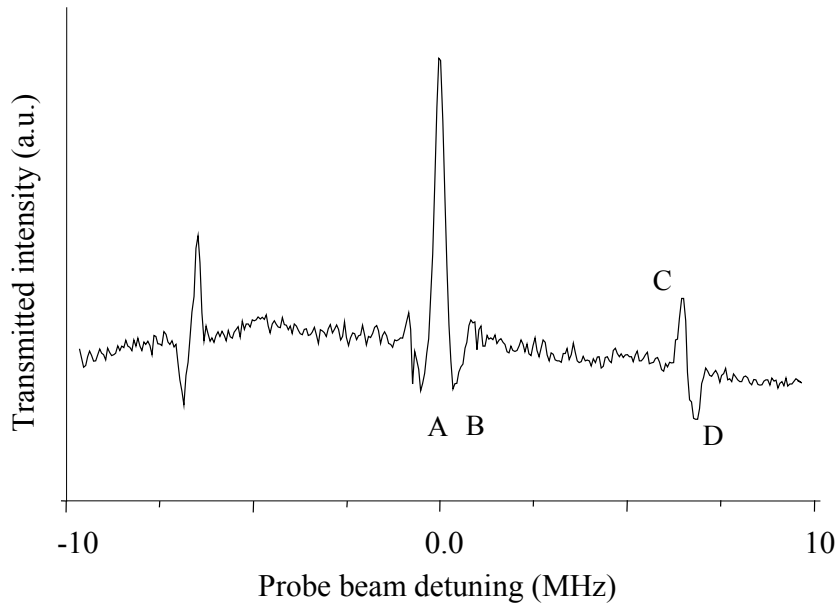

Figure 5 hole-burning transmission spectrum for measuring level splitting at $B_y = 0$. The 0.45T magnetic field is oriented along $[\bar{1}\bar{1}1]$. The light beam is polarized along [111] so that sites 1, 3 and 5 are selected. Holes and anti-holes from site 1 fall away from the explored spectral window. The holes and anti-holes B, C and D are located at distance $\left(\Delta_g - \Delta_e\right)$, $\Delta_e$, and $\Delta_g$ respectively from burning position A.

Ions in site 1 also interact with the light beam. The transmission increase at burning position A partly reflects the depletion of site 1 ions. However, the side holes and anti-holes related to site 1 are far away from the analyzed spectral window. Optical pumping of site 1 ions out of the spectral region of interest offers a convenient way to focus on ions in sites 3 and 5 in their optimized magnetic field direction.

## 3.4 Measurement of the gyromagnetic tensor y-component

As already mentioned, in order to burn persistent holes we must slant the magnetic field with respect to the site local axes. Otherwise the optical transition cannot flip the nuclear spin. Therefore to perform $\gamma_y$ measurement we consider two situations where the applied magnetic field lies either in plane $xOy$ or in plane $yOz$ in local frames.

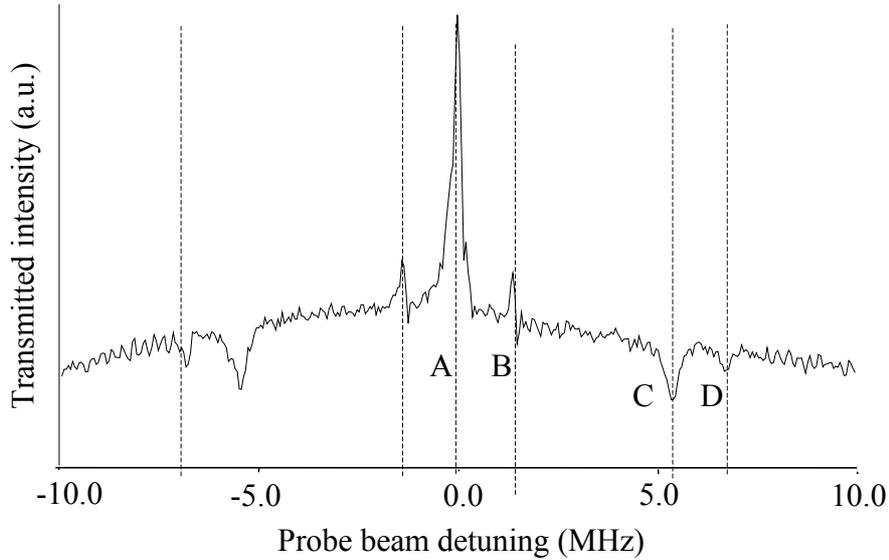

Figure 6 hole-burning transmission spectrum for measuring $\gamma_y$. The 0.024T magnetic field is oriented along [001]. The light beam, polarized along the same direction, probes the equivalent sites 3, 4, 5 and 6. The holes and anti-holes B, C and D are located at distance $\Delta_e$, $\left( \Delta_g - \Delta_e \right)$, and $\Delta_g$ respectively from burning position A.

Since $\gamma_y$ is expected to be much larger than $\gamma_x$ and $\gamma_z$, we reduce the magnetic field to 0.024T in order to keep the level splitting size within reach of the acousto-optic scanning range. In the first experiment the magnetic field is directed along the vertical axis of the crystal cell [001]. The light beam is polarized along the same direction [001]. This way the four sites 3, 4, 5 and 6 are equivalent with respect to interaction with both the magnetic field and the light beam. In the four site frames the applied magnetic field is directed along the diagonal of plane $xOy$. The sites 1 and 2 are cross-polarized with the light beam. The spectrum in Figure 6 represents a 750μs-long scan over a 20MHz interval after a 30ms waiting time. Burning is observed at the central position A. The hole B is ascribed to a transition to the upper level and should be located at distance $\Delta_e$ from the burning position. The anti-holes C and D are expected to be located at $\Delta_g - \Delta_e$ and $\Delta_g$ respectively from the burning position. From the data we deduce the following splitting values:

$$\Delta_e^{[001]} = 60 \pm 2\, MHz / Tesla \,, \; \Delta_g^{[001]} = 285 \pm 2\, MHz / Tesla \qquad (23)$$

where the connection with $\gamma_x$ and $\gamma_y$ reads as:

$$\Delta^{[001]} = \sqrt{\left( \gamma_x^2 + \gamma_y^2 \right) / 2} \qquad (24)$$

Anti-hole C appears to be broader than hole B and anti-hole D. This difference can be ascribed to saturation effect. Indeed B and D both result from burning along a forbidden transition, with the same pumping rate $P(\omega - \omega_0) R / (1 + R_1)$, while C results from burning along an allowed transition with pumping rate $P(\omega - \omega_0) R_1 / (1 + R_1)$. Saturation can affect C without broadening B and D if $R$ is significantly smaller than $R_1$. According to Eqs. (7), (8) and (12), the branching ratio $R$ reads as:

$$R_{[001]} = \frac{1}{4} \left( r_e - r_g \right)^2 \qquad (25)$$

In the second experiment the 0.024T-magnetic field is directed along the diagonal of the crystal cell [111]. The light beam is polarized along the same direction [111]. This way the sites 1, 3, and 5 are equivalent and sites 2, 4, and 6 are cross-polarized with the light beam. In the relevant site local frames the applied field lies in plane $yOz$ at angle $\sin^{-1}(1/\sqrt{3})$ from direction $Oy$. The sample is probed in 750µs over a 17MHz-wide interval after a 10ms waiting time. In the transmission spectrum displayed in Figure 7, the absence of saturation and of burning by the probe suggest the temperature is similar to that in previous experiment. The only visible anti-holes are assigned to the burning transition features located at $\Delta^{(g)} - \Delta^{(e)}$ from the burning line. Their distance of 260MHz/T from burning frequency is rather close to the 230MHz/T value reported by Macfarlane [14]. The absence of features at $\Delta^{(g)}$ and $\Delta^{(e)}$ shows that $R << \Gamma_0 / P(0)$ and the presence of an anti-hole at $\Delta^{(g)} - \Delta^{(e)}$ means that $R << R_1$. The branching ratio $R$ reads as:

$$R_{[111]} = \frac{1}{8} \left( s_e - s_g \right)^2 \tag{26}$$

where $s_i = \gamma_z^{(i)} / \gamma_y^{(i)}$ for $i = e$ or $g$.

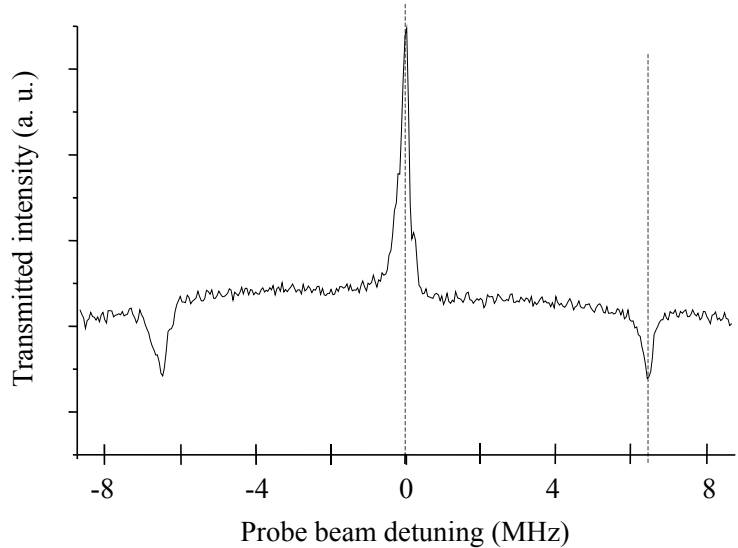

Figure 7 hole-burning transmission spectrum for measuring $\gamma_y$. The 0.024T magnetic field is oriented along [111]. The light beam, polarized along the same direction, probes the equivalent sites 1, 3, 5. The only observed anti-holes are located at distance $\Delta_g - \Delta_e$ from the burning position.

Since spectra in Figure 7 and Figure 6 have been obtained in similar temperature conditions, we are led to:

$$R_{[111]} / \Gamma_0 << R_{[001]} / \Gamma_0 \tag{27}$$

that is to say, according to Eqs. (25) and (26):

$$\left(s^{(e)} - s^{(g)}\right)^2 << \left(r^{(e)} - r^{(g)}\right)^2 \qquad (28)$$

Because of the expected smallness of the ground state transverse factors $r^{(g)}$ and $s^{(g)}$, the relation entails that: $\left(\gamma_x^{(e)}\right)^2 >> \left(\gamma_z^{(e)}\right)^2$, which is consistent with the theoretical predictions [24].

To observe more holes and anti-holes we have to immerse the sample in superfluid helium at 1.4K. The corresponding spectrum is displayed in Figure 8. Burning is observed at the central position A. The hole C is ascribed to transition to the upper level and should be located at distance $\Delta_e$ from the burning position. The anti-holes E and G are expected to be located at $\Delta_g - \Delta_e$ and $\Delta_g$ respectively from the burning position.

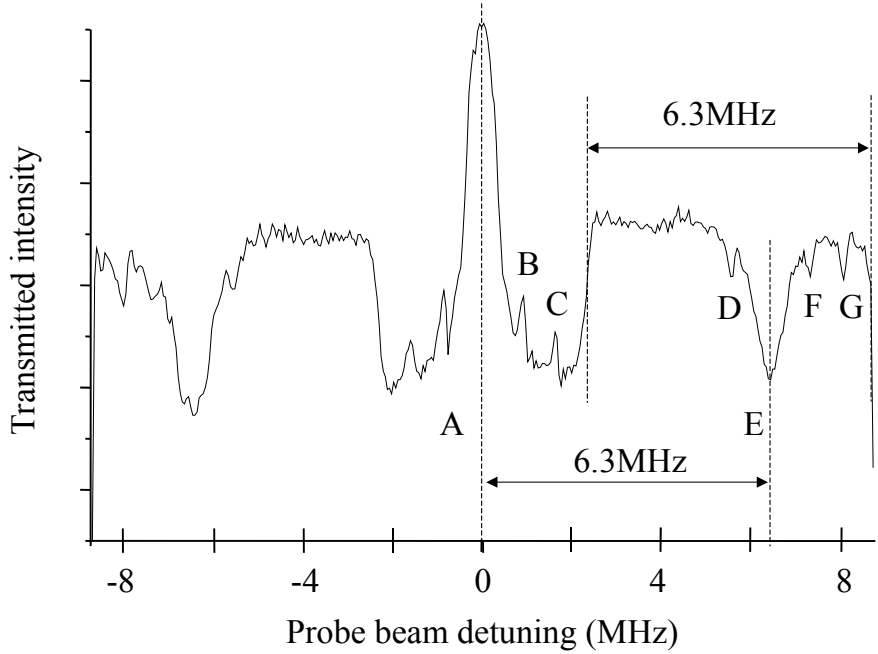

Figure 8 hole-burning transmission spectrum for measuring $\gamma_y$. Same configuration as in Figure 7 but the sample is immersed in superfluid helium. The holes and anti-holes C, E and G are located at distance $\Delta_e$, $\Delta_g - \Delta_e$, and $\Delta_g$ respectively from burning position A. The additional structures are detailed in the text.

The hole B and the anti-holes D and F are ascribed to a nearby high power (300kW) broadcasting transmitter at 864kHz that modulates the laser through the locking loop, adding two side bands to the burning field. Therefore holes B and C are different in nature. Hole B corresponds to read out of a hole burnt by a side band at the very frequency at which burning took place. On the contrary, hole C corresponds to read out of the central hole at a frequency different from the burning frequency. In the latter case the readout field excites a transition to the other upper sub-level. The symmetric character of the modulated burning field can be observed at the anti-hole position: D and F reflect burning by the side bands. They are evenly spaced from E. Since it appears on a single side of E, the anti-hole G is not related to a burning side band.

In addition one can observe sharp cliffs located at 6.3MHz from the edges of the probed interval. We verified that the distance from the edges does not change when we vary the window size. This figure coincides with the distance $\Delta_g - \Delta_e$ of anti-hole E from the central frequency. This feature very likely reflects the burning action of the probe beam. When ions located at distance $> \left( \Delta_g - \Delta_e \right)$ from the edges are pumped to the other nuclear sublevel by the probe, they can later be pumped back to their initial position, since their both allowed transition frequencies are contained within the probed interval. On the contrary, ions located at distance $< \left( \Delta_g - \Delta_e \right)$ from the edges cannot be taken back to their initial position by the probe if the transition frequency to their other ground state sublevel lies outside the probed window. In summary, to burn holes through the forbidden transition excitation one has to cool the sample down to a temperature where the small population transfer caused by the probe is accumulated during the very long nuclear spin lifetime.

From the data in Figure 8 we deduce the following splitting values:

$$\Delta_e^{[111]} = 67 \pm 2 MHz / Tesla , \ \Delta_g^{[111]} = 329 \pm 2 MHz / Tesla \tag{29}$$

where the connection with $\gamma_y$ and $\gamma_z$ reads as:

$$\Delta^{[111]} = \sqrt{\left( 2\gamma_y^2 + \gamma_z^2 \right) / 3} \tag{30}$$

To get $\gamma_y$ one can just insert the splitting data into the following combination of Eqs. (22), (24), and (30): $\gamma_y = \sqrt{\left(\Delta^{[001]}\right)^2 + \frac{3}{4}\left[\left(\Delta^{[111]}\right)^2 - \left(\Delta^{[\bar{1}\bar{1}1]}\right)^2\right]}$. The error bar on $\gamma_y$ then amounts to ~10MHz/Tesla. We can get better precision by exploiting the relative size of the $\gamma's$. Indeed according to data (21) and (29), and to our finding that $\left(\gamma_x^{(e)}\right)^2 >> \left(\gamma_z^{(e)}\right)^2$, $\gamma_z$ contribution to $\Delta^{[111]}$ can be neglected in both ground state and excited state, which leads to:

$$\gamma_y^{(g)} = 403 \pm 3 MHz / Tesla \ , \ \gamma_y^{(e)} = 82 \pm 3 MHz / Tesla \qquad (31)$$

One easily checks that these values are consistent with $\Delta^{[001]}$ data. The experimental data show reasonable quantitative agreement with theoretical predictions as illustrated in **Table 2**. The predicted qualitative features we summarized in section 2.2 are confirmed by experiment, except the smallness of anisotropy disparity between ground and excited states in plane $xOz$. Indeed we failed to separately measure $\gamma_x$ and $\gamma_z$.

| | | lower level $^3H_6$ | upper level $^3H_4$ |
|---|---|---|---|
| experiment | $\gamma_y$ | $403 \pm 3$ MHz/T | $82 \pm 3$ MHz/T |
| | $\gamma_x / \gamma_y$ | <0.05 | $0.21 \pm 0.01$ |
| | $\sqrt{2\gamma_x^2 + \gamma_z^2} / \gamma_y$ | 0.066 | 0.297 |
| theory | $\gamma_y$ | 560 MHz/T | 75 MHz/T |
| | $\gamma_x / \gamma_y$ | 0.033 | 0.3 |
| | $\gamma_z / \gamma_y$ | 0.02 | 0.080 |
| | $\sqrt{2\gamma_x^2 + \gamma_z^2} / \gamma_y$ | 0.051 | 0.432 |

Table 2 experimental data and theoretical predictions.

Then, calculating the lower boundary to the optimal branching ratio with the help of Eq. (18) and of data (21) and (31), one finally gets:

$$R_{\max} \geq 0.13 \pm 0.01 \qquad (32)$$

When same intensity laser beams excite both transitions along the $\Lambda$ legs, the Rabi frequency ratio is given by the branching ratio square root that appears to be better than 0.36. The optimal branching ratio value is obtained when the applied field is oriented at angle:

$$\delta\Theta_0 = 2\theta_0 / \sqrt{3} = 2\sqrt{\Delta_g \Delta_e} / \sqrt{3\gamma_y^{(g)}\gamma_y^{(e)}} = 5.4 \pm 0.2° \qquad (33)$$

from direction $[\bar{1}\,\bar{1}1]$. Measurements of $R_{max}$ and $\delta\Theta_0$ represent the main result of this work.

## 4 Conclusion

From the experimental data we conclude that a three-level $\Lambda$ system can operate efficiently in $Tm^{3+}$:YAG, the optical transitions along the $\Lambda$ legs exhibiting a Rabi frequency ratio better than 0.36 when excited at the same light intensity. In addition, despite of the rather complex site structure, the equivalent ions contributing to the $\Lambda$ system operation can represent as much as 2/3 of the available optical density. Experimental results agree quite well with theory. The predicted large gyromagnetic tensor anisotropy along the local site axis $Oy$ has been observed, together with the large anisotropy disparity between ground and excited states. The physical origin of these features should be investigated in depth.

In the present work the transition probability ratio has been deduced from splitting measurements. The next steps will involve the direct measurement of this ratio together with the measurement of the lower level superposition state lifetime. The latter parameter is of critical importance for optical storage in the spin state.

## ACKNOWLEDGMENT

This work was carried out under the ESQUIRE project supported by the IST-FET program of the EC.

**APPENDIX**

## Lower boundary to the branching ratio optimal value

It is possible to express a lower boundary to the branching ratio optimal value in terms of the level splitting at $B_y = 0$. Let the applied magnetic field coordinates be expressed as:

$$\vec{B} = B \begin{vmatrix} \cos\theta\cos\varphi \\ \sin\theta \\ \cos\theta\sin\varphi \end{vmatrix} \tag{A.1}$$

in the site frame. Then the effective field unit vector reads as:

$$\hat{B}_{eff} = \frac{1}{D} \begin{vmatrix} r \\ \tan\theta/\cos\varphi \\ s\tan\varphi \end{vmatrix} \tag{A.2}$$

where $r = \gamma_x/\gamma_y$, $s = \gamma_z/\gamma_y$ and $D = \sqrt{r^2 + s^2(\tan\varphi)^2 + (\tan\theta/\cos\varphi)^2}$. Then one can calculate the angle of the effective field unit vectors in ground and excited level with the help of their cross product. One obtains:

$$\left\| \hat{B}_{eff}^{(e)} \times \hat{B}_{eff}^{(g)} \right\|^2 = \left(\sin\alpha_{eff}\right)^2 =$$
$$\frac{1}{D_e^2 D_g^2}\left\{ (\tan\theta/\cos\varphi)^2 \left( \left[(s_g - s_e)\tan\varphi\right]^2 + \left[(r_g - r_e)\right]^2 \right) + \left[(r_e s_g - r_g s_e)\tan\varphi\right]^2 \right\} \tag{A.3}$$

The cross product vanishes when $B_y = 0$ and the gyromagnetic tensor anisotropy in plane $xOz$ is identical in both electronic states, i.e. $r_e s_g = r_g s_e$. Keeping equal $xOz$ anisotropy in both electronic states and increasing $B_y$, one observes that the cross product departs from zero and reaches maximum value at:

$$\tan\theta_0 = \left[ \sqrt{r_e^2 + (s_e\tan\varphi)^2} \sqrt{r_g^2 + (s_g\tan\varphi)^2} \right]^{1/2} \cos\varphi \tag{A.4}$$

The maximum position is slightly different when $xOz$ anisotropy is not the same in upper and ground electronic states, but a tractable lower boundary value of the cross product is readily calculated at $\theta_0$. At $\theta_0$ the quantity $D_e^2 D_g^2$ reads as:

$$D_e^2 D_g^2 = \sqrt{r_e^2 + \left(s_e \tan\varphi\right)^2} \sqrt{r_g^2 + \left(s_g \tan\varphi\right)^2} \left[\sqrt{r_e^2 + \left(s_e \tan\varphi\right)^2} + \sqrt{r_g^2 + \left(s_g \tan\varphi\right)^2}\right]^2 \qquad (A.5)$$

We also notice that:

$$\left[\left(s_g - s_e\right)\tan\varphi\right]^2 + \left[\left(r_g - r_e\right)\right]^2 =$$
$$\left[\sqrt{r_e^2 + \left(s_e \tan\varphi\right)^2} - \sqrt{r_g^2 + \left(s_g \tan\varphi\right)^2}\right]^2 + 2\frac{\left[\left(r_e s_g - r_g s_e\right)\tan\varphi\right]^2}{\sqrt{r_e^2 + \left(s_e \tan\varphi\right)^2}\sqrt{r_g^2 + \left(s_g \tan\varphi\right)^2} + r_e r_g + s_g s_e \left(\tan\varphi\right)^2}$$
$$(A.6)$$

Finally one obtains:

$$\left(\sin\alpha_{eff}\right)^2 = \left(\frac{\Delta_g / \gamma_y^{(g)} - \Delta_e / \gamma_y^{(e)}}{\Delta_g / \gamma_y^{(g)} + \Delta_e / \gamma_y^{(e)}}\right)^2 + A \times C \times F \qquad (A.7)$$

where $\Delta_e$ and $\Delta_g$ represent the splitting in excited and ground state at $\theta = 0$, $A$ stands for the anisotropy disparity coefficient in plane $xOz$:

$$A = \left(\frac{r_e s_g - r_g s_e}{r_e s_g + r_g s_e}\right)^2 \qquad (A.8)$$

and:

$$C = \left[\frac{\left(r_e s_g + r_g s_e\right)\tan\varphi}{D_e D_g}\right]^2 = \frac{1}{\delta_e \delta_g}\left[\frac{\left(t_g + t_e\right)\tan\varphi}{\delta_e \sqrt{r_e / r_g} + \delta_g \sqrt{r_g / r_e}}\right]^2 \qquad (A.9)$$

$$F = \left[1 + 2\frac{\delta_e \delta_g}{1 + \delta_e \delta_g + t_e t_g \left(\tan\varphi\right)^2}\right] \qquad (A.10)$$

where:

$$t_i = \gamma_z^{(i)} / \gamma_x^{(i)}, \ \delta_i = \sqrt{1 + \left(t_i \tan\varphi\right)^2} \qquad (A.11)$$

The cross product expression reduces to

$$\sin \alpha_{eff} = \frac{\left| \Delta_g / \gamma_y^{(g)} - \Delta_e / \gamma_y^{(e)} \right|}{\Delta_g / \gamma_y^{(g)} + \Delta_e / \gamma_y^{(e)}} \qquad (A.12)$$

at $\varphi = 0$ and $\varphi = \pi / 2$. More generally, since $A$, $C$ and $F$ are positive numbers we may write:

$$\left( \sin \alpha_{eff} \right)_{\max} \geq \frac{\Delta_g / \gamma_y^{(g)} - \Delta_e / \gamma_y^{(e)}}{\Delta_g / \gamma_y^{(g)} + \Delta_e / \gamma_y^{(e)}} \qquad (A.13)$$

$$R_{\max} \geq \left( \frac{\sqrt{\Delta_g / \gamma_y^{(g)}} - \sqrt{\Delta_e / \gamma_y^{(e)}}}{\sqrt{\Delta_g / \gamma_y^{(g)}} + \sqrt{\Delta_e / \gamma_y^{(e)}}} \right)^2 \qquad (A.14)$$

In terms of the level splitting and of $\gamma_y$, the local frame direction that optimizes the branching ratio reads as:

$$\tan \theta_0 = \sqrt{\Delta_g \Delta_e} / \sqrt{\gamma_y^{(g)} \gamma_y^{(e)}} \qquad (A.15)$$

The factors $C$ and $F$ reduce to simple expressions under the conditions predicted by theory in **Table 2**, namely:

$$\left( t_i \tan \varphi \right)^2 \ll 1, \ r_g \ll r_e \qquad (A.16)$$

Then :

$$C \simeq \frac{r_g}{r_e} \left[ \left( t_g + t_e \right) \tan \varphi \right]^2 \ll 1 \ \text{and} \ F \simeq 2 \qquad (A.17)$$

It appears that the $C$ factor attenuates the effect of anisotropy disparity in the plane *xOz*.

# REFERENCES

[1] J. Frenkel, Phys. Rev. **37**, 17-44; 1276-1294 (1931)

[2] J. Hald, J. L. Sørensen, C. Schori, and E. S. Polzik, Phys. Rev. Lett. **83**, 1319 (1999).

[3] A. Kuzmich, L. Mandel, and N.P. Bigelow, Phys. Rev. Lett. **85**, 1594 (2000).

[4] B. Julsgaard, A. Kozhekin, and E. S. Polzik, Nature **413**, 400 (2001).

[5] B. Julsgaard, J. Sherson, J. I. Cirac, J. Fiurášek, and E. S. Polzik Nature **432**, 482(2004);

[6] C. H. van der Wal, M. D. Eisaman, A. André, R. L. Walsworth, D. F. Phillips, A. S. Zibrov, and M. D. Lukin, Science **301**, 196 (2003).

[7] A. Kuzmich, W.P. Bowen, A.D. Boozer, A. Boca, C.W. Chou, L.-M. Duan and H.J. Kimble, Nature **423**, 731 (2003).

[8] D. N. Matsukevich and A. Kuzmich, Science **306** (2004) 663

[9] M. D. Lukin, Rev. Mod. Phys. **75**, 457 (2003),

[10] E. Fraval, M.J. Sellars, and J.J. Longdell , Phys. Rev. Lett. **92**, 077601 (2004)

[11] E. Fraval, M.J. Sellars, and J.J. Longdell, Phys. Rev. Lett. **95,** 030506 (2005)

[12] A. V. Turukhin, V.S. Sudarshanam, M.S. Shahriar, J.A. Musser, B.S. Ham, P.R. Hemmer, Phys. Rev. Lett. 88, 023602 (2002)

[13] J.J. Longdell, E. Fraval, M. J. Sellars, N. B. Manson, *Stopped light with storage times greater than one second using EIT in solids* (2005), quantph/0506233

[14] R M. Macfarlane, J. Lumin. **100**, 1 (2002).

[15] M. Mitsunaga, N. Uesugi, K. Sugiyama, Opt. Lett. **18**, 1256 (1993).

[16] M. J. Sellars, R. S. Meltzer, P. T. H. Fisk, N. B. Manson, J. Opt. Soc. Am. B **11**, 1468 (1994)

[17] R. Klieber, A. Michalowski, R. Neuhaus, and D. Suter, Phys Rev B **67**, 184103 (2003)

[18] A. Schoof, J. Grünert, S. Ritter, A. Hemmerich, Opt. Lett. **26** (2001) 1562, and references therein.

[19]    Z. Cole, T. Böttger, R. Krishna Mohan, R. Reibel, W. R. Babbitt, R. L. Cone, and K. D. Merkel, Appl. Phys. Lett. **81** (2002) 3525.

[20]    V. Lavielle, I. Lorgeré, and J.-L. Le Gouët, S. Tonda and D. Dolfi, Opt. Lett. **28** (2003) 384.

[21]    R. M. Macfarlane, Opt. Lett. **18,** 1958 (1993).

[22]    N. Ohlsson, M. Nilsson, S. Kröll and R. Krishna Mohan, Opt. Lett. **28**, 450 (2003).

[23]    P. Goldner and O. Guillot-Noël, Mol. Phys. **102**, 1185 (2004).

[24]    O. Guillot-Noël, Ph. Goldner, E. Antic-Fidancev and J. L. Le Gouët, Phys. Rev. B **71** 174409 (2005).

[25]    A. Abragam and B.Bleaney, "Electron Paramagnetic Resonance of Transition Ions", Oxford University Press, 1970.

[26]    A.A. Kaplyanskii, R.M. Macfarlane, "Spectroscopy of solids containing rare earth ions," pp 86-89, Elsevier Science Publishers B.V., Amsterdam, 1987.

[27]    M. A. Teplov, Zh. Eksp. Teor. Fiz. **53**, 1510 (1967) [Sov. Phys. JETP **26** 872 (1968)].

[28]    Y. Sun, G. M. Wang, R.L. Cone, R.W. Equall, M. J. M. Leask, Phys. Rev. B **62**, 15443-15451 (2000).

[29]    V. Crozatier, F. de Sèze, L. Haals, F. Bretenaker, I. Lorgeré and J.-L. Le Gouët Opt. Commun. **241,** 203 (2004).